\documentclass[11pt]{article}
\usepackage{amsmath}
\usepackage{amsfonts}

\def\be{\begin{equation}}
\def\ee{\end{equation}}
\def\bea{\begin{eqnarray}}
\def\eea{\end{eqnarray}}

\def\hR{\hat{R}}
\def\hg{\hat{g}}
\def\hM{\hat{M}}

\def\hT{\hat{T}}

\def\scri{\mathcal{I}}
\def\hM{\hat{M}}

\def\hg{\hat{g}}
\def\hM{\hat{M}}

\def\hT{\hat{T}}

\def\eqq{\hat{=}}
\def\ra{\stackrel{\circ}{\alpha}}
\def\rb{\stackrel{\circ}{\beta}}
\begin{document}

\title{Penrose's quasi-local mass for asymptotically anti-de Sitter space-times}
\author{Ron Kelly\\
Type-set and annotated by Paul Tod}

\maketitle
\begin{abstract}
Penrose's quasi-local mass construction is carried through for two-surfaces at infinity in asymptotically anti-de Sitter space-times. A modification of the Witten argument is given to prove a 
positivity property of the resulting conserved quantities.

\medskip

\emph{This work formed part of Ron Kelly's Oxford D.Phil. thesis, and the first person pronoun refers to him. It appeared in hand-written form as `Asymptotically anti-de Sitter space-times' in 
Twistor Newsletter 20 (1985) pp11-23,\footnote{{\bf{PT}}: available at http://people.maths.ox.ac.uk/lmason/Tn/TN1-25.} but is appearing type-set for the first time here. 
Footnotes marked ${\bf{PT}}$ have been added for this version by Paul Tod, in the hope of making this work available to a wider audience.}

\end{abstract}

\section{The Angular Momentum Twistor}

I will first briefly review the work of Ashtekar and Magnon \cite{am}: a space-time $(\hM,\hg_{ab})$ is said to be \emph{asymptotically anti-de Sitter} 
if there exists a manifold $M$ with boundary $\partial M$ and metric $g_{ab}$ and a diffeomorphism from $\hM$ to $M-\partial M$ such that
\begin{enumerate}
 \item there is a smooth real-valued function $\Omega$ on $M$ such that $g_{ab}=\Omega^2\hg_{ab}$ on $\hM$;
 \item $\scri:=\partial M$ is topologically $S^2\times\mathbb{R}$ and $\Omega=0$ on $\scri$;
 \item $\hg_{ab}$ satisfies
 \be\label{e1}
 \hR_{ab}-\frac12\hR\hg_{ab}+\lambda\hg_{ab}=-8\pi G\hT_{ab}\ee
 with $\lambda<0$ and where $\Omega^{-4}\hT_a^b$ has a smooth limit on $\scri$;
 \item write $B_{ab}$ for the magnetic part of the Weyl tensor of $M$ then $B'_{ab}:=\Omega^{-1}B_{ab}$ vanishes on $\scri$.
 \end{enumerate}
In fact in \cite{am} it was assumed instead in item 3 that $\Omega^{-3}\hT_a^b$ has a limit on $\scri$; the choice made here preserves the conservation equation.

It will be convenient to introduce the notation $\eqq$ to mean `equals at $\scri$', so that for example $\Omega\eqq 0$ and $B'_{ab}\eqq 0$. 
 Examination of the Bianchi identities shows that
 \begin{itemize}
  \item[(i)] with $s_a=\partial_a\Omega$ we have $s^as_a\eqq\lambda/3$ so that $\scri$ is time-like; 
  \item[(ii)] by modifying $\Omega$ one can require $\nabla_as_b\eqq 0$;
  \item[(iii)] $C_{abcd}\eqq 0$;
  \item[(iv)] condition 4 above is equivalent to the vanishing of the Cotton-York tensor of $\scri$, so that $\scri$ is conformally flat. It is also equivalent to the condition 
  \be\label{e2}
  D_{[a}V_{b]c}\eqq 0,\ee
  where $D_a$ is the intrinsic covariant derivative of $\scri$ and 
  \[V_{ab}=\Phi_{ab}-\Lambda g_{ab}-E_{ab},\]
  where in turn $\Phi_{ab}=-\frac12(R_{ab}-\frac14Rg_{ab})$, $\Lambda=R/24$ and $E_{ab}$ is the electric part of the Weyl tensor.
 \end{itemize}
From the discussion in \cite{t1}, conditions (iii) and (iv) in the second list are sufficient to conclude that $\scri$ may be embedded in conformally-flat space-time with the same 
first and second fundamental forms, and thus that 3-surface twistors exist on $\scri$\footnote{{\bf{PT}}: Three-surface twistors are defined on a three-surface $\Sigma$ with normal $n^a$ 
in a space-time by taking those components of the twistor equation
\[\nabla_{AA'}\omega_B=-i\epsilon_{AB}\pi_{A'}\] for which the derivative is projected tangent to the three-surface. Writing $D_{AB}=n_{(A}^{\;\;A'}\nabla_{B)A'}$, the 3-surface twistor equation is
\[D_{AB}\omega_{C}=-i\pi_{(A}\epsilon_{B)C}\mbox{  with  }\pi_A=n_A^{\;\;A'}\pi_{A'}.\]}.

We may therefore think of $\scri$ as the conformal infinity of anti-de Sitter space which 
can be embedded in the Einstein static cylinder as the product of the time axis with the equatorial 2-sphere of the 3-sphere cross-sections. In the usual way, the conformal group of $\scri$ is the anti-de Sitter group $O(2,3)$ 
and there are ten linearly independent conformal Killing vectors of $\scri$\footnote{{\bf{PT}}: So these generate conformal symmetries of $\scri$.}. Hawking \cite{h1} has shown that condition 4 in the first list is equivalent to the assumption that 
gravitational radiation satisfies a reflective boundary condition at $\scri$.

Following \cite{am}, given a cross-section $C$ of $\scri$ and a conformal Killing vector $\xi^a$ of $\scri$ one defines a conserved quantity
\be\label{e3}
Q_\xi[C]:=-\frac{1}{8\pi G}\oint E'_{ab}\xi^adS^b,\ee
where $E'_{ab}=\Omega^{-1}E_{ab}$. This expression is conformally invariant and has a flux $F_\xi[\Delta]$ through a region $\Delta$ of $\scri$ bounded by two cross-sections given by
\be\label{e4}
F_\xi[\Delta]=\int_\Delta\left({\mbox{Lim}}_{\Omega\rightarrow 0}\Omega^{-4}\hT_a^b\right)s^a\xi_bd\Sigma.\ee
If there is no matter near to $\scri$ then the flux vanishes.

\medskip

The Schwarzschild--anti-de Sitter metric can be written
\[ds^2=\left(1-\frac{2GM}{r}+a^2r^2\right)dt^2-\left(1-\frac{2GM}{r}+a^2r^2\right)^{-1}dr^2-r^2(d\theta^2+\sin^2\theta d\phi^2),\]
with $\lambda=-3a^2$. Choosing $\xi^a=\frac{1}{a}\frac{\partial}{\partial t}$, which with $\Omega=r^{-1}$ is unit at $\scri$ in the rescaled metric, we obtain
\[Q_\xi[C]=M\]
on any cross-section $C$ of $\scri$.

To show that the expression (\ref{e3}) coincides with Penrose's expression for quasi-local mass and angular momentum \cite{p1} (modified for asymptotically anti-de Sitter by the subtraction of the 
cosmological constant term) we first need the following Lemma:

\medskip

\noindent{\bf{Lemma}}

\noindent Suppose $\Sigma$ is a hypersurface in a conformally-flat space-time, with unit normal $\zeta_a$ and $\nabla_a\zeta_b$ zero at $\Sigma$; then the vector field
\be\label{e6}\xi^a:=\omega^A\zeta^{A'}_{\;B}\omega^B\ee
is a (null) conformal Killing vector at $\Sigma$ if and only if
\be\label{e5}\nabla_{A'(A}\omega_{B)}=0\ee
at $\Sigma$.\footnote{{\bf{PT}}: The field $\omega^A$ here is a 3-surface twistor on $\scri$ so (\ref{e5}) might better be written as $D_{(AB}\omega_{C)}=0$}.
 
The proof is simply by substitution of the solution of the twistor equation (\ref{e5}) in a conformally-flat space-time. For a general conformal Killing vector one has a sum of terms like (\ref{e6}). 

\medskip

On $\scri$, with unit normal $\zeta^a=s^a/a$, take $\xi^a=2i\omega^A\zeta^{A'}_{\;B}\omega^B$ then with a cross-section $C$ of $\scri$ with unit normal $t^a$ tangent to $\scri$ we have
\[Q_\xi[C]:=-\frac{1}{8\pi G}\oint_C E'_{ab}\xi^adS^b=-\frac{1}{8\pi G}\oint_C 2\phi_{ABCD}\zeta^C_{\;A'}\zeta^D_{\;B'}2i\omega^A\zeta^{A'}_{\;B}t^{BB'}dS\]\[=-\frac{i}{4\pi G}\oint_C\phi_{ABCD}\omega^A\omega^Bo^C\iota^DdS,\]
where $\phi_{ABCD}=\Omega^{-1}\psi_{ABCD}$ and $\psi_{ABCD}$ is the Weyl spinor. This is now recognisable as Penrose's expression for the quasi-local mass and angular momentum \cite{p1}.\footnote{{\bf{PT}}: Here there 
has been a calculation, that $\zeta^C_{\;A'}\zeta^D_{\;B'}\zeta^{A'}_{\;B}t^{BB'}=o^{(C}\iota^{D)}$, where $o^A,\iota^A$ are spinors representing the out and 
ingoing null normals to the cross-section $C$, normalised to have $o_A\iota^A=1$. This formula establishes the connection between Penrose's quasi-local kinematic quantities and the Ashtekar-Magnon charges.}

We shall choose our momentum and angular momentum conformal Killing vectors by embedding $\scri$ as the boundary of anti-de Sitter space in the Einstein cylinder and then restricting solutions of the twistor 
equation in conformally-flat space-time to it. This process depends on how we embed Minkowski space-time into the Einstein static cylinder with respect to anti-de Sitter space-time. 
We choose to do this symmetrically\footnote{{\bf{PT}}: This means the following: take the Minkowski metric to be
\[g_M=dT^2-dR^2-R^2(d\theta^2+\sin^2\theta d\phi^2),\]
then this is conformal to the metric of the Einstein static cylinder written as
\[g_E=dt^2-dr^2-\sin^2r(d\theta^2+\sin^2\theta d\phi^2),\]
with $T\pm R=\tan(t\pm r)/2$ so that the worldlines $r=0$ and $R=0$ coincide. The anti-de Sitter metric with $a=1$ can then be written
\[g_{adS}=(\sec^2r)g_E,\]
and occupies half the Einstein static cylinder, with $r\leq\pi/2$.}.

With respect to a constant spinor basis $(\alpha^A,\beta^A)$ in Minkowski space the solution to the twistor equation (see eg \cite{ht}) is given by

\[\omega^A=\Omega^A-ix^{AA'}\pi_{A'},\]
where
\[\Omega^A=\Omega^0\alpha^A+\Omega^1\beta^A,\;\;\pi_{A'}=\pi^{0'}\bar{\alpha}_{A'}+\pi^{1'}\bar{\beta}_{A'}\]
and $\Omega^0,\Omega^1,\pi^{0'},\pi^{1'}$ are complex constants. Also $x^{AA'}$ is the position vector in Minkowski space in Cartesians.

Take the Einstein cylinder to have metric
\[ds^2=dt^2-dr^2-\sin^2r(d\theta^2+\sin^2\theta d\phi^2),\]
and introduce the null tetrad
\[\ell=\frac{1}{\sqrt{2}}\left(\frac{\partial}{\partial t}+\frac{\partial}{\partial r}\right),\;\;n=\frac{1}{\sqrt{2}}\left(\frac{\partial}{\partial t}-\frac{\partial}{\partial r}\right),\;\;
m=\frac{1}{\sqrt{2}\sin r}\left(\frac{\partial}{\partial \theta}-\frac{i}{\sin\theta}\frac{\partial}{\partial \phi}\right).\]
This tetrad implicitly defines a spinor dyad $(o^A,\iota^A)$ in the usual way and expanded in this dyad the solution of the twistor equation  is
\[\omega^A=\omega^0o^A+\omega^1\iota^A\]
with
\[\omega^0=\sqrt{2}\cos(\frac12(t+r))\left(\Omega^0e^{-i\phi/2}\cos(\theta/2)+\Omega^1e^{i\phi/2}\sin(\theta/2)\right)\]\[+i\sin(\frac12(t+r))\left(-\pi^{0'}e^{i\phi/2}\sin(\theta/2)+\pi^{1'}e^{-i\phi/2}\cos(\theta/2)\right),\]
\be\label{om1}\omega^1=\sqrt{2}\cos(\frac12(t-r))\left(-\Omega^0e^{-i\phi/2}\sin(\theta/2)+\Omega^1e^{i\phi/2}\cos(\theta/2)\right)\ee\[+i\sin(\frac12(t-r))\left(\pi^{0'}e^{i\phi/2}\cos(\theta/2)+\pi^{1'}e^{-i\phi/2}\sin(\theta/2)\right).\]
Since the twistor equation is conformally invariant, this is also a solution of the twistor equation on the Einstein cylinder.

From these expressions we may calculate $Q_\xi[C]$ with the conformal Killing vector
\[\xi^a=2i\zeta^{A'}_{\;B}\omega^{(A}\tilde{\omega}^{B)}\]
and the cross-section $C$ in $r=\pi/2$ as
\be\label{amt1}Q_\xi[C]=-\frac{i}{4\pi G}\oint_C\phi_{ABCD}\omega^A\tilde{\omega}^BdS^{CD}=A_{\alpha\beta}Z^\alpha\tilde{Z}^\beta,\ee
with
\[Z^\alpha=(\Omega^A,\pi_{A'}),\;\;\tilde{Z}^\alpha=(\tilde{\Omega}^A,\tilde{\pi}_{A'})\]
and
\[A_{\alpha\beta}=
\left(\begin{array}{cc}2\Phi_{AB} &P_A^{\;B'}\\ P^{A'}_{\;B} &\Phi^{A'B'}\end{array}\right)
\]
In particular the 4 components of momentum $P^{\bf{a}}$  are given by
\[Q_\xi[C]:=-\frac{1}{8\pi G}\oint E'_{ab}\xi^adS^b\]
with $\xi^a$ one of $(\gamma^a,\frac12(\eta^a+\bar{\eta}^a),\frac{1}{2i}(\eta^a-\bar{\eta}^a),\beta^a)$ where
\[\gamma=\frac{\partial}{\partial t},\;\eta=e^{i\phi}(\sin t\sin\theta\frac{\partial}{\partial t}-\cos t\cos\theta\frac{\partial}{\partial \theta}-i\cos t\frac{\partial}{\partial \phi}),\]
\[\beta=\cos\theta\sin t\frac{\partial}{\partial t}+\sin\theta\cos t\frac{\partial}{\partial \theta}.\]
For the case of Schwarzschild--anti-de Sitter $P^{\bf{a}}=(M,0,0,0)$ and $\Phi_{AB}=0$ and we recover $M$ as the mass.

The angular momentum automatically obeys the Hermiticity property
\be\label{hp1}A_{\alpha\beta}I^{\beta\gamma}=\overline{A_{\gamma\beta}I^{\beta\alpha}}\ee
with respect to the infinity twistor\footnote{{\bf{PT}}: Remember that, in this part of the discussion, $a=1$; otherwise $a^2$ appears in (\ref{inf1}).}
\be\label{inf1}I^{\alpha\beta}=\left(\begin{array}{cc}-\frac12\epsilon^{AB}&0\\0&\epsilon_{A'B'}\end{array}\right).\ee
\section{A positive energy theorem}
We first need to look at the work of Gibbons et al \cite{g1}. Consider a space-like hypersurface $\Sigma$ in an asymptotically anti-de Sitter space-time (possibly with an inner boundary on a past 
or future apparent horizon ${\mathcal{H}}$) which asymptotically approaches the $t=0$ cross-section of $\scri$. Define a \emph{supercovariant} derivative on a 4-spinor $(\alpha^A,\beta^{A'})$ by 
\begin{eqnarray}
\hat{\nabla}_{MM'}\alpha_A&=&\nabla_{MM'}\alpha_A+\frac{a}{\sqrt{2}}\epsilon_{MA}\beta_{M'}\label{sccs}\\
\hat{\nabla}_{MM'}\beta_{A'}&=&\nabla_{MM'}\beta_{A'}+\frac{a}{\sqrt{2}}\epsilon_{M'A'}\alpha_M.\nonumber
\end{eqnarray}
Now suppose that $D_a$ is the projection into $\Sigma$ of $\nabla_a$ and introduce the \emph{supercovariant Witten equation} on $\Sigma$:
\begin{eqnarray*}
\hat{D}_{AA'}\alpha^A&=&D_{AA'}\alpha^A+\frac{3a}{2\sqrt{2}}\beta_{A'}=0\\
\hat{D}_{AA'}\beta^{A'}&=&D_{AA'}\beta^{A'}+\frac{3a}{2\sqrt{2}}\alpha_{A}=0.
\end{eqnarray*}
Then one can show that
\be\label{id1}
-D_m\left(t^{AA'}(\bar{\alpha}_{A'}\hat{D}^m+\bar{\beta}_A\hat{D}^m\beta_{A'})\right)=-t^{AB'}(\hat{D}_m\alpha_A\overline{\hat{D}^m\alpha_B}+\hat{D}_m\beta_{B'}\overline{\hat{D}^m\beta_{A'}})+4\pi GT_{ab}t^a\xi^b,
\ee
where $\xi^a=\alpha^A\bar{\alpha}^{A'}+\beta^{A'}\bar{\beta}^A$. Thus if $T_{ab}$ satisfies the Dominant Energy Condition, then the RHS is non-negative. We may choose the spinors $\alpha^A,\beta^{A'}$ to obey boundary 
conditions on an inner apparent horizon ${\mathcal{H}}$ such that
\[t^{AA'}(\bar{\alpha}_A'\hat{D}^m\alpha_A+\bar{\beta}_A\hat{D}^m\beta_{A'})N_m=0 \mbox{  at }{\mathcal{H}},\]
where $N_m$ is the normal to $\mathcal{H}$ lying in $\Sigma$.

If we use Green's Theorem on the identity (\ref{id1}) we therefore obtain
\be\label{id2}
-\oint_{t=0}t^{AA'}(\bar{\alpha}_A'\hat{D}^m\alpha_A+\bar{\beta}_A\hat{D}^m\beta_{A'})s_mdS\geq 0.\ee
This integral is finite providing that
\[\hat{D}_m\alpha_A\rightarrow 0,\;\;\hat{D}_m\beta_{A'}\rightarrow 0\mbox{ in the limit at  }\scri.\]
In \cite{g1} it was proposed that the boundary term could be written as
\be\label{id3}4\pi G\left(P^{(5)AA'}(\ra_A\overline{\ra}_{A'}+\rb_{A'}\overline{\rb}_A)-\lambda^{AB}\ra_A\overline{\rb}^B-\overline{\lambda}^{A'B'}\overline{\alpha}_{A'}\overline{\beta}_{B'}\right)\ee
(credited as a private communication from D.Z.Freedman) where $\ra_A,\rb_{A'}$ are the limits on $\scri$ of the supercovariantly constant spinors $\alpha_A,\beta_{A'}$ (after division by a suitable power of the conformal factor).

For Schwarzschild--anti-de Sitter the component $P^{(5)0}:=\frac{1}{\sqrt{2}}(P^{(5)00'}+P^{(5)11'})$ is a (constant, positive) multiple of the mass parameter $M$. In general, the inequality (\ref{id2}) 
implies $P^{(5)0}\geq 0$ so that, if we identify this term with the mass, as was done in \cite{g1}, then we have shown that the mass is positive. But, as shall be described, if we evaluate the integral in (\ref{id2}) 
explicitly we obtain Penrose's expression $Q_\xi[C]$ and then we find that (\ref{id2}) contains much more information about $A_{\alpha\beta}$.

We first write the metric in terms of coordinates $(u,s,\theta,\phi)$ as\footnote{{\bf{PT}}: The leading terms in this expression correspond to the metric $g_{adS}$ of footnote 6 but with general $a$.}
\be\label{met1}
g=\frac{1}{\sin^2s}(du^2-\frac{2}{a}duds-\cos^2s(d\theta^2+\sin^2\theta d\phi^2))+O(s^{-1})
\ee
where the $O(s^{-1})$ terms do not contain $ds^2$. Here the constant $u$ surfaces are outgoing null hypersurfaces meeting $\scri$ which is located at $s=0$, $(x^2,x^3)=(\theta,\phi)$ label the null generators 
of the constant $u$ surfaces, and $s$ is a parameter (not affine) on each null generator. We shall choose a null tetrad as follows:
\[\ell_adx^a=\frac{du}{\sin s},\mbox{   so that   }\ell^a\partial_a=-a\sin s\frac{\partial}{\partial s},\]
$n^a$ is the ingoing null normal to the 2-spheres of constant $u$ and $s$, and $m^a$ is a complex null tangent to these 2-spheres, chosen so that the NP spin coefficient $\epsilon$ is 
real\footnote{{\bf{PT}}: This can be accomplished by rotating
 $m^a$ in the 2-plane tangent to the chosen 2-spheres.}. We 
thus have
\begin{eqnarray}
 \ell^a\partial_a&=&-a\sin s\frac{\partial}{\partial s},\\
 n^a\partial_a&=&\sin s\left(\frac{\partial}{\partial u}+aU\frac{\partial}{\partial s}+X^k\frac{\partial}{\partial x^k}\right),\\
 m^a\partial_a&=&\frac{a}{\sqrt{2}}\xi^k\frac{\partial}{\partial x^k},
\end{eqnarray}
for real $U,X^k$ and complex $\xi^k$. For anti-de Sitter space-time
\[U=\frac12,\;\;X^k=0,\;\;\xi^2=\tan s,\;\;\xi^3=-i\tan s/\sin\theta,\]
so that these are the values at $\scri$ in general. Also from the assumption that $\Omega^{-4}\hat{T}_{ab}$ is finite at $\scri$ we have the asymptotic behaviour of the NP curvature quantities:
\[\Phi_{ij}=O(s^4),\;\;\Lambda=-\lambda/6=O(s^4),\;\;\psi_i=O(s^3).\]
We may therefore solve for the NP spin coefficients as power series expansions in $s$ and then we may calculate $\alpha_A,\beta_{A'}$ similarly. We find
\[\alpha^A=s^{-1/2}\sum_0^\infty\alpha_i^As^i,\;\;\beta^{A'}=s^{-1/2}\sum_0^\infty\beta_i^{A'}s^i,\]
with $\beta_0^{A'}=\sqrt{2}\alpha_0^As_A^{A'}$ (this equality is an identity in the rescaled space-time since $s^a\eqq\frac12\ell^a-n^a$).

The integral on the LHS of (\ref{id2}) becomes 
\[-\frac{\sqrt{2}}{a^3}\oint_{t=0,s=0}\left(\psi_1^{(3)}\alpha^0_0+\psi_2^{(3)}(\alpha^1_0\bar{\beta}^0_0+ \alpha^0_0\bar{\beta}^1_0)+    \psi_3^{(3)}\alpha^1_0\bar{\beta}^1_0\right)dS_0\]
\[=-\frac{i}{a^3}\oint_{t=0,s=0}\left(\psi_1^{(3)}\omega^0\tilde{\omega}^0+\psi_2^{(3)}(\omega^1\tilde{\omega}^0+\omega^0\tilde{\omega}^1)+\psi_3^{(3)}\omega^1\tilde{\omega}^1\right)dS_0\]
\[=\frac{4\pi G}{a^3}A_{\alpha\beta}Z^\alpha\tilde{Z}^\beta,\]
where $\psi_i^{(3)}$ means the $O(s^3)$ term in $\psi_i$, and $\omega^A\eqq\alpha^A_0,\tilde{\omega}^A\eqq-i\sqrt{2}\bar{\beta}^A_0$ which are found to be 2-surface twistors on the $t=0$ 
cross-section of $\scri$. But in (\ref{om1}) we have expressions for $\omega^0,\omega^1$ in terms of $(\Omega^A,\pi_{A'})$ at $t=0$, and the relation between $\alpha_0^A$ and $\bar{\beta}^{A'}_0$ 
gives us that 
\[\tilde{Z}_\alpha=2I^{\alpha\beta}\bar{Z}_\beta,\]
using (\ref{inf1}). We have therefore shown, by (\ref{id2}), that
\be\label{pos1}A_{\alpha\beta}I^{\beta\gamma}Z^\alpha\bar{Z}_\gamma\geq 0.\ee
In particular this implies that $P^a$ is time-like and future-pointing\footnote{{\bf{PT}}: Kelly didn't make the point explicitly but this result also provides \emph{rigidity} in the sense that if there exists a 
$Z^\alpha$ giving zero in (\ref{pos1}) then the space-time is exactly anti-de Sitter at least near $\Sigma$. This follows from (\ref{id1}): Dominant Energy implies that $T_{ab}$ vanishes at $\Sigma$, 
and therefore everywhere; and one can evolve $(\alpha^A,\beta^{A'})$ from $\Sigma$ to obtain a solution of (\ref{sccs}) in the space-time, which forces the Weyl tensor to vanish.}. From the angular momentum twistor $A_{\alpha\beta}$ we may calculate the associated mass $m_P$ as
\[m_P^2=-\frac12 A_{\alpha\beta}\bar{A}^{\alpha\beta}=P_aP^a-\Phi_{AB}\Phi^{AB}-\bar{\Phi}_{A'B'}\bar{\Phi}^{A'B'}.\]
Inequality (\ref{pos1}) implies that $m_P^2$ is non-negative and also provides a further inequality relating components of $P^a$ and $\Phi_{AB}$\footnote{{\bf{PT}}: In this connection, see \cite{cmt}.}. An alternative 
definition of mass is
\[m_D^4=4\,\mbox{det}A_{\alpha\beta}=4\epsilon^{\alpha\beta\gamma\delta}A_{\alpha 1}A_{\beta 2}A_{\gamma 3}A_{\delta 4},\]
and $m_D^4$ is also non-negative by virtue of (\ref{pos1}). When $\Phi_{AB}=0$, we have $P^aP_a=m_P^2=m_D^2$ but in general the masses are different.

\medskip

{\bf{Acknowledgements}}

I would like to thank Dr Paul Tod for considerable assistance in this work and Professor Penrose and William Shaw for useful discussions. I hope to publish a more detailed version of this work in the near future.


\end{document}